\documentstyle[twocolumn,prl,aps,epsf]{revtex} 
\begin{document}
\def\abstract#1{\begin{center}{\large Abstract}\end{center} \par #1}

\title{\bf Do naked singularities generically occur in generalized 
theories of gravity?} 
\author{Kengo Maeda \thanks{Electronic address: maeda@th.phys.titech.ac.jp}, 
Takashi Torii \thanks{Electronic address: torii@th.phys.titech.ac.jp}, 
and Makoto Narita \thanks{Electronic address: narita@se.rikkyo.ac.jp}} 
\address{${}^{\mbox{\rm *\dag}}$
Department of Physics, Tokyo institute of Technology, Oh-Okayama, 
Meguro, Tokyo 152, Japan}
\address{${}^{\ddag}$
Department of Physics, Rikkyo University, Nishi-Ikebukuro, Toshima, 
Tokyo 171, Japan}
\maketitle
\abstract{A new mechanism for causing naked singularities is 
found in an effective superstring theory. 
We investigate the gravitational collapse in a spherically 
symmetric Einstein-Maxwell-dilaton system in the presence of a pure 
cosmological constant ``potential", where the system has no static black 
hole solution. We show that once gravitational collapse occurs in 
the system, naked singularities necessarily appear in the sense 
that the field equations break down in the domain of outer 
communications. This suggests that in generalized theories of 
gravity, the non-minimally coupled fields generically cause naked 
singularities in the process of gravitational collapse if the system 
has no static or stationary black hole solution. }\\

The singularity theorem\cite{HP} states that the occurrence of 
singularities is inevitable under some physical conditions in 
general relativity. There are two notable scenarios where 
singularities may appear in our universe. 
One is the initial singularity at the 
birth of the universe and the other is the final stage of 
gravitational collapse. In the latter case we believe that an 
event horizon is formed which encloses all occurring 
singularities as the collapse proceeds, following the cosmic censorship 
hypothesis (CCH)~\cite{P}. CCH is classified into two types, 
the weak cosmic censorship hypothesis (WCCH) and the strong one (SCCH). 
WCCH says that observers at an infinity should not see singularities 
while SCCH says that no observer should see them and the whole 
region of a space-time can be uniquely determined by initial 
regular data. Mathematically, SCCH is equivalent to the 
statement that no Cauchy horizon can be formed in a physical 
gravitational collapse.

Recently many elegant results~\cite{B} suggest that SCCH holds 
for charged and/or rotating black holes due to the destruction 
of the Cauchy horizon by the mass inflation phenomenon. The general 
proof of CCH is, however, far from complete and many counter 
examples have been found in the framework of general 
relativity~\cite{W}. A new approach has been considered in the 
context of generalized theories of gravity. 
Gibbons and Maeda~\cite{GM} discovered the static black hole 
solutions in the Einstein-Maxwell-dilaton system, which comes 
from an effective superstring theory, and later 
Garfinkel, Horowitz and Strominger~\cite{GHS} showed that 
the inner (Cauchy) horizon in the Reissner-Nordstr{\"o}m solution 
is replaced by a spacelike singularity. 
This suggests that the occurrence of the inner horizon is not 
generic and hence SCCH holds if we take the effect of string 
theory into account. On the other hand Horne and Horowitz~\cite{H} 
obtained the opposite result that extremal electrically 
charged non-static black hole solutions in the presence of 
a central charge have timelike singularities. 
It seems, however, not to be a counter example of SCCH in 
the sense that such solutions have no regular initial 
spacelike hypersurface because of the central 
singularities. Thus, the following question naturally arises. 
Does CCH really hold in generalized theories of gravity? 
There is not much evidence deciding the matter yet because 
the above results do not take any physical process of the 
gravitational collapse from initial regular data into account. 

Although a large number of studies have been made on finding 
static/stationary black hole solution and investigating their 
thermodynamical properties and geometrical stability and so on 
in generalized theories of gravity, only few studies have so far 
been made on investigating gravitational collapse. 
We should not overlook that the gravitational collapse in 
generalized theories of gravity will be much different from 
that in general relativity. In general relativity, the black hole
no-hair conjecture states that the matter fields are swallowed 
by a black hole and a space-time asymptotically approaches the 
electrovacuum Kerr-Newman solution after gravitational collapse. 
On the other hand, the above aspects are not necessarily satisfied 
in generalized theories of gravity because the dilaton field 
couples to the curvature and/or other matter fields.

In this letter we investigate the gravitational collapse in the 
spherically symmetric Einstein-Maxwell-dilaton system analytically 
in the presence of a pure positive cosmological constant, which 
corresponds to $g_1=0$ in the Liouville-type potential~\cite{PW}. 
This system is interesting because it has been proved that 
no static black hole solution exists for spherically symmetric 
space-times~\cite{PT}. Such property seems to be general in the sense 
that there exists no asymptotically flat, asymptotically de Sitter, 
or asymptotically anti-de Sitter solution in the arbitrary 
exponential dilaton potential~\cite{PW}. Furthermore, it is known 
that there is a critical mass below which no asymptotically 
flat black hole solution exists in the system which includes the 
Gauss-Bonnet term which was neglected in the former works~\cite{PK,TYM}. 
In these systems the following three interpretations are 
available: (i) the matter fields do not cause the gravitational 
collapse but escape to infinity without forming a black hole event 
horizon~(BEH); (ii) the space-time does not approach a 
stationary space-time and the matter fields oscillate forever or, 
(iii) naked singularities are necessarily formed. 
If the case (i) is true for all initial data, we should find 
that the strong cosmic no-hair conjecture holds in the present system. 
By investigating numerically, however, we confirmed that this is 
not the case and some initial data lead to the gravitational 
collapse, namely, a trapped surface forms~\cite{MSTN}. 
This result is consistent with the fact that the dilaton field 
satisfies the dominant energy condition in the Einstein frame. 
Then it is enough to consider only the cases (ii) and (iii). 
The case (ii) seems incompatible with our naive expectation that 
after gravitational collapse, the space-time settles down 
to an asymptotically stationary space-time with a black hole. 
If the case (iii) is true, we must say that CCH is violated even 
in generalized theories of gravity. 

Using the double null coordinates, we will show that the field 
equations necessarily break down in the domain of outer communications 
or at the BEH, once the gravitational collapse has occurred. 
It is worth to note that ``break down" does not imply a  
coordinate singularity because our coordinates are, probably, the 
most extended coordinates in spherically symmetric space-times. 
It is also supported by numerical calculations which show 
the formation of naked singularities in the system~\cite{MSTN}. Thus, 
our result suggests that naked singularities generically occur in the 
gravitational collapse of spherically symmetric systems. 

\begin{figure}[htbp]
\centerline{\epsfsize=6.0cm \epsfbox{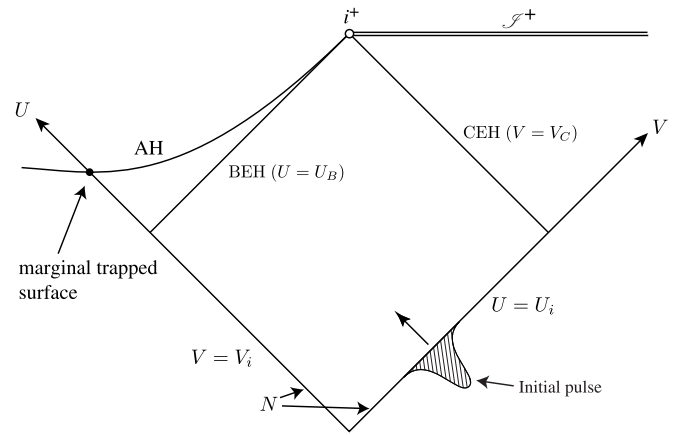}} 
\caption{A Penrose diagram in asymptotically de Sitter space-time. 
$N$ and {\bf AH} are the characteristic hypersurface and the 
apparent horizon of the black hole, respectively. }
\protect
\label{UV-eps}
\end{figure}

The low-energy effective action (Einstein frame) of string theory as a 
dilaton model is 
\begin{eqnarray}
\label{act}
S=\int d^4 x \sqrt{-g}\,
[-R+2(\nabla\phi)^2+e^{-2\phi} F^2+2\Lambda], 
\end{eqnarray}
where $R$ is the Ricci scalar, $\phi$ is a massless dilaton field, 
$\Lambda$ is a positive constant, and 
$F_{\mu\nu}$ is the field strength of the Maxwell field.
The double-null coordinates for a spherically symmetric space-time are
\begin{eqnarray}
\label{eq-du}
ds^2=-2e^{-\lambda}(U, V)dU dV + R(U, V)^2 d\Omega, 
\end{eqnarray}
where $\partial_U$ and $\partial_V$ are future-ingoing and outgoing null 
geodesics, respectively. 
The Maxwell equation is automatically satisfied for a 
purely magnetic Maxwell field 
$F=Q\sin\theta d\theta\wedge d\phi\, (F^2=2Q^2/R^4)$, where 
$Q$ is the magnetic charge.
An electrically charged solution is obtained by a duality rotation 
from the magnetically charged one~\cite{GM,GHS}. 
Therefore we shall consider only a purely magnetic case in this letter. 
The dynamical field equations are 
\begin{eqnarray}
\label{eq-la}
\lambda_{,UV}-\frac{2 R_{,UV}}{R}=
2\phi_{,U}\phi_{,V} + e^{-\lambda}
\left(\frac{Q^2 e^{-2\phi}}{R^4}-\Lambda \right), 
\end{eqnarray}
\begin{eqnarray}
\label{eq-R}
R_{,UV}+\frac{R_{,U}R_{,V}}{R}
=-\frac{e^{-\lambda}}{2R}
\left(1-\frac{Q^2 e^{-2\phi}}{R^2}
-\Lambda R^2 \right),
\end{eqnarray}
\begin{eqnarray}
\label{eq-ddi}
2R^3 (R_{,U}\phi_{,V}+R{{\phi}_{,UV}}
+R_{,V}\phi_{,U})
=Q^2 e^{-2\phi-\lambda},
\end{eqnarray}
where $A_{,a}$ is a partial derivative of $A$ with respect to $a$. 
The constraint equations are 
\begin{eqnarray}
\label{eq-ru}
{R}_{,UU}+{\lambda}_{,U}{R}_{,U}
=-({\phi}_{,U})^2 R,
\end{eqnarray}
\begin{eqnarray}
\label{eq-rv}
{R}_{,VV}+{\lambda}_{,V}{R}_{,V}
=-({\phi}_{,V})^2 R.
\end{eqnarray}
We consider the evolution of the field equations with initial 
regular data on a null characteristic hypersurface whose boundary is a 
closed future-trapped surface (see Fig.~\ref{UV-eps}) because we are 
interested in the gravitational collapse. 
As shown in Theorem~1 of Ref.~\cite{SK}, the trapped surface causes the 
formation of a BEH~($U=U_B$ null hypersurface)
if there are no singularities observed from ${\cal I}^+$. 
For this initial value problem, we will show the following theorem. \\
{\bf Theorem}\\
{\it Let us consider the dynamical evolution of the model~(\ref{act}) 
in a spherically symmetric space-time or equivalently for the 
equations~(\ref{eq-la})-(\ref{eq-rv}) with initial data on the 
characteristic null hypersurface $N$. Then, there is $U_1(\le U_B)$ 
such that the system of equations breaks down at $U=U_1$}. 

We shall show the above theorem by contradiction below. First, we will 
consider the asymptotic behavior of field functions near the cosmological 
event horizon (CEH), which is defined as a past Cauchy horizon 
$H^{-}({\cal I}^+)$ when a BEH exists (see Ref.~\cite{M}).
It is convenient to rescale the coordinate $U$ such that $U$ is an affine 
parameter of a null geodesic of the CEH, i.e., $\lambda$ is constant 
along the CEH. 
Hereafter we use a character $u(=f(U))$ instead of $U$ for 
parameterization to avoid confusion. 
Under such coordinates, Eq.~(\ref{eq-ru}) on the CEH is 
\begin{eqnarray} 
\label{eq-ruc}
-R_{,uu}=(\phi_{,u})^2 R.
\end{eqnarray}
If there are no singularities in 
$\overline{J^{-}({\cal I}^+)}\cap J^+(N)$, 
the null geodesic generators of the BEH and the CEH are future complete. 
Then, by the non-decreasing area law for the CEH and by 
existence of an upper bound of the area~\cite{M}, 
$\lim_{u \to \infty} R=C_1$ (hereafter, $C_i(i=1,2,..)$ means a 
positive constant), and hence future asymptotic 
behavior of $R_{,u}$ on the CEH is represented as follows~\cite{OT}, 
\begin{eqnarray}
\label{eq-dR}
R_{,u}\sim C_2 u^{-\alpha-1} \quad (\alpha >0). 
\end{eqnarray}
Then, by Eq. (\ref{eq-ruc})
\begin{eqnarray}
\label{eq-dph}
\phi_{,u}\,\sim u^{-\alpha/2-1},
\end{eqnarray}
and hence 
$\lim_{u \to \infty} \phi(u) =\mbox{const.}$ 
We shall obtain the asymptotic value of 
$R_{,V}$ on the CEH by solving Eq.~(\ref{eq-R}) as 
\begin{eqnarray}
R_{,V} &=& \left(\int_{u_i}^u \frac{K R}{R_i}du+ R_{,V}|_i \right)
\frac{R_i}{R},
\end{eqnarray}
where $R_{,V}|_i,\,R_i$ are initial values of $R_{,V},\,R$ on $u=u_i$, 
respectively and $K=-(1-Q^2e^{-2\phi}/R^2-\Lambda R^2 )/2R e^\lambda$. $K$ 
must approach a positive constant $K_\infty$ as $u \to\infty$ 
because $R,\,\phi\to \mbox{const.}$ and the expansion 
$\theta_{+}\equiv R_{,V}/R$ of each outgoing null geodesic is 
positive (if it were negative, the area element $d\Omega$ would 
become $0$ along the outgoing null geodesics). 
Then, the asymptotic value of $R_{,V}$ is 
\begin{eqnarray}
\label{eq-R'}
R_{,V}\,\sim\,K_\infty u>0.
\end{eqnarray}
We shall also obtain the asymptotic value of $\phi_{,V}$ on the CEH 
by solving Eq.~(\ref{eq-ddi}) with the following solution 
\begin{eqnarray}
\label{eq-sph'}
\phi_{,V} &=& \left(\int_{u_i}^u \frac{H R}{R_i}du+ \phi_{,V}|_i \right)
\frac{R_i}{R},
\end{eqnarray}
where $\phi_{,V}|_i$ is an initial value of $\phi_{,V}$ on $u=u_i$ and 
$H = ( Q^2 e^{-2\phi-\lambda}-2R^3 R_{,V}\phi_{,u})/2 R^4$. $H$ 
approaches a positive constant $H_\infty$ as $u \to\infty$ because 
$R_{,V}\phi_{,u}\sim u^{-\alpha/2} \to 0$ by Eqs.~(\ref{eq-dph}) 
and (\ref{eq-R'}). 
The asymptotic value of $\phi_{,V}$ is 
\begin{eqnarray}
\label{eq-ph'}
\phi_{,V}\sim H_\infty u>0.
\end{eqnarray}

Next, let us consider an infinitesimally small neighborhood~${\cal U}_C$ 
of the CEH which contains a timelike hypersurface $T_C$ 
such that $T_C$ is in the past of the CEH by $\epsilon(>0)$, 
where $\epsilon$ is a fixed affine parameter 
distance along the outgoing 
null geodesic intersecting the CEH~\cite{Affin}. We denote 
each point of the intersection of $T_C$ and $u=\mbox{const.}$~hypersurface 
by $p(u)$. 
By Eq.~(\ref{eq-ddi}), the solution of $h\equiv\phi_{,u}$ 
along each $u=\mbox{const.}$ is 
\begin{eqnarray}
\label{eq-pu}
h &=& \left(-\int_{V}^{V_C} \frac{LR}{R_C}dV+ h_C \right) \frac{R_C}{R},
\end{eqnarray}
where $R_C$ and $h_C$ are values of $R$ and $h$ at the CEH~($V=V_C$), 
respectively, and 
$L = (Q^2 e^{-2\phi-\lambda}-2R^3 R_{,u}\phi_{,V})/2R^4$. 
Since 
$(R_{,u}\phi_{,V}/R)|_{V_C}\sim u^{-\alpha}\to 0$ for 
large values of $u$, $L\sim L_\infty>0$ asymptotically. 
Differentiating $h$ by $V$ in Eq.~(\ref{eq-pu}), 
\begin{eqnarray}
\label{eq-hv}
h_{,V}=L+
\frac{R_C R_{,V}}{R^2}
\left(\int_{V}^{V_C} \frac{LR}{R_C}dV
-h_C \right).
\end{eqnarray}
By the relation $\epsilon=dr\sim u dV $ for large $u$~\cite{MTN}, 
$h_{,V}\sim L_{\infty}+O(\epsilon) + O(u^{-\alpha/2})$. 
Since $\epsilon$ is an arbitrary small value, asymptotically 
$h_{,V}>0$ on the $u=\mbox{const.}$ null segment 
$[V|_{p(u)}, V_C]\subset {\cal U}_C$. 

In the next step we investigate the behavior of $\phi,\phi_{,V}$ and 
$R$ on BEH, just like the CEH case. 
We rescale $V$ into $v$ such that $v$ is an affine parameter and 
$\lambda=\mbox{const.}$ on the BEH, while we leave $U$ unchanged. 
Since the area of the BEH is non-decreasing and also has an 
upper bound as shown in~\cite{SK,M}, 
$ \lim_{v \to \infty} R=C_3$, and 
%\label{eq-drv}
$R_{,v}\sim C_4 v^{-\beta-1}~(\beta>0)$. 
%\end{eqnarray}
This means that by Eq.~(\ref{eq-rv}) $\phi_{,v}\sim v^{-\beta/2-1}$ as 
obtained in the CEH case. 
By replacing $u$ by $v$ in the argument of the CEH case and 
solving Eqs.~(\ref{eq-R}) and (\ref{eq-ddi}), the asymptotic values of 
$\phi_{,U}$ and $R_{,U}$ become 
%\begin{eqnarray}
%\label{eq-drv}
$R_{,U} \sim v$ and $\phi_{,U} \sim C_5 v$, respectively. 

\begin{figure}[htbp]
\centerline{\epsfsize=6.0 cm \epsfbox{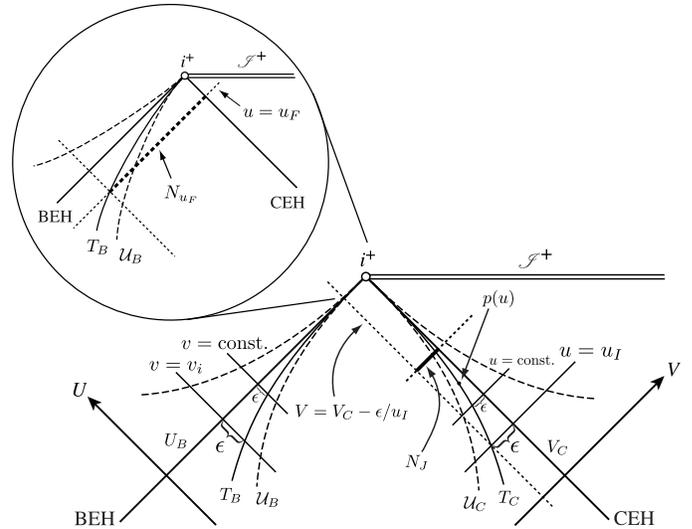}} 
\caption{Timelike hypersurfaces $T_B,\,T_C$ are displayed in the 
neighborhoods ${\cal U}_B$, ${\cal U}_C$ of BEH and CEH, respectively. 
A null segment $N_J$ is displayed by a thick line. $\epsilon$ is a 
fixed affine parameter distance along outgoing and 
ingoing null geodesics.}
\protect
\label{Show-eps}
\end{figure}

Hence the first and third terms in the l.h.s. of the dilaton 
field Eq.~(\ref{eq-ddi}) are negligible asymptotically and 
\begin{eqnarray}
\label{eq-C}
\lim_{v \to \infty}
k_{,U}=
\lim_{v \to \infty}
\phi_{,vU}=C_6>0.
\end{eqnarray}
We define $k\equiv\phi_{,v}$. 
Consider an infinitesimally small neighborhood ${\cal U}_B$ of the BEH. 
There is a small $\epsilon$ such that the timelike hypersurface $T_B$ which is 
in the past of the BEH by a fixed affine parameter distance 
$\epsilon$ of ingoing null geodesics intersecting the BEH is 
contained in ${\cal U}_B$. 
By the relation between the affine parameter and 
$dU$,\,i.e., $dU \propto \epsilon/v$, $k$ on $T_B$ is asymptotically 
\begin{eqnarray}
k|_{T_B} &\sim&
k|_{BEH} + {k_{,U}|_{BEH}}~(-dU) \nonumber \\ &\sim& k|_{BEH} - C_6
\,\epsilon\, v^{-1} \sim (v^{-\beta/2} - C_6\,\epsilon)v^{-1}.  
%\sim -C_6\,\epsilon\, v^{-1}.
\end{eqnarray}
This indicates that $\phi_{,V}$ is negative on $T_B$ for large values 
$v(>v_1)$. 
Now, consider each null segment $N_{u}(u\ge u_I) :[V_C-\epsilon/u_I,\,V_C]$ 
where $h_{,V}>0$ on $N_{u_I}$~(see Fig.~\ref{Show-eps}). If one takes $u_I$ 
large enough, $N_{u}$ intersects $T_B(v>v_1)$ at $u=u_F$. Let us take a 
sequence of $N_{u_J}(J=1,2,...,L+1)$~($L$ is a natural number large enough), 
where $\delta u=(u_F-u_I)/L$ and $u_J=u_I+(J-1) \delta u$.
Assume that $\phi_{,V}>0$ on $N_J(\equiv N_{u_J})$, for a moment, then 
$h_{,V}>0$ on $N_{J}$ in the same way as $h_{,V}>0$ on 
$[V|_{p(u)}, V_C]$~\cite{MK}. Hence we can show that 
$\phi_{,V}|_{N_{J+1}}\cong
(\phi_{,V}|_{N_J}+h_{,V}|_{N_J}\delta u)>0$. 
On the other hand, $\phi_{,V}>0$ on $u=u_I$ by Eq.~(\ref{eq-ph'}), 
hence $\phi_{,V}>0$ for each $u_J$ by induction. 
This is a contradiction because as we showed $\phi_{,V}<0$ 
on $T_B(u=u_F)$.~\qquad~$\Box$  

It is worth to comment that the possibility of the formation of null 
singularities on BEH, continuing to $i^+$ is not excluded from our theorem. 
In this case the BEH is a singular null hypersurface even 
if WCCH holds~(SCCH is, probably, violated). 

Our theorem says that the case (ii) is not true, as we expected. When we 
consider the dyon solution or a rotating space-time, 
non-trivial 3-rank anti-symmetric tensor fields inevitably 
appear. However, we expect that such fields do not change our 
result drastically. This implies that if there are 
no static/stationary black hole solutions, naked singularities generically 
occur in the process of gravitational collapse in generalized 
theories of gravity. 

Since the model under consideration is the low energy limit of 
string theory, we cannot exclude the possibility that higher 
curvature terms and higher order $\alpha'$ corrections, where $\alpha'$ is 
the inverse string tension, may prevent the space-time from causing 
naked singularities. However, the theory is useful near the Planck scale. 
Thus, we can at least say that {\it string theory predicts existence of 
space-time points with very high curvature in the domain of outer 
communications}. 
This picture is quite different from that of general relativity 
because it is strongly believed that space-time points with high curvature 
do not appear generically in the outer region of a black hole. 

\section*{Acknowledgments}
We would like to thank Akio Hosoya and Hideki Ishihara and Kei-ichi Maeda 
for useful discussions, and Christian Baraldo for his critical reading 
of our letter. 
The work was supported partially by the Grant-in-Aid for Scientific Research 
Fund of the Ministry of Education, Science, Sports and Culture~(K.M and T.T), 
by the Grant-in-Aid for JSPS~(No. 199605200~(K.M) and No. 199704162~(T.T)).

%%%%%%%%%%%%%%%%%%%%%%%%%%%%%%%%%%%%%%%%%%%%%%%%%%%%%%%%%%%%%%% 
%%%%%%%%%%%%%%%%%%%%%%%%%%%%%%%%%%%%%%%%%%%%%%%%%%%%%%%%%%%%%%%%% 

%%%%%%%%%%%%%%%%%%%%%%%%%%%%%%%%%%%%%%%%%%%%%%%%%%%%%%%%%%%%%%% 
%%%%%%%%%%%%%%%%%%%%%%%%%%%%%%%%%%%%%%%%%%%%%%%%%%%%%%%%%%%%%%%%% 


\begin{thebibliography}{99}
\bibitem{HP}
See S. W. Hawking and G. F. R. Ellis,
{\it The large scale structure of space time} 
(Cambridge University Press, Cambridge, 1973). 

\bibitem{P}R. Penrose, Riv. Nuovo Cimento {\bf 1}~(1969)~252. 

\bibitem{B}P.~R.~Brady, C.~M.~Chambers,
Phys.~Rev.~D {\bf 51} (1995) 4177.\,
L. M. Burko, Phys. Rev. Lett. {\bf 79}~(1997)~4958.\, P.~R.~Brady, I.~G.~Moss, and 
R.~C.~Myers, Phys. Rev. Lett. {\bf 80}~(1998)~3432.

\bibitem{W}A review is written by R. M. Wald, gr-qc/9710068. 

\bibitem{GM} G. W. Gibbons and K. Maeda, Nucl. Phys. B298, (1988)~741.

\bibitem{GHS}D. Garfinkle, G. T. Horowitz and A. Strominger, Phys. Rev. D~{\bf 43} 
(1991) 3140. 

\bibitem{H}J. H. Horne and G. T. Horowitz, Phys. Rev. D~{\bf 48} (1993) R5457.

\bibitem{PW}S. J. Poletti, and D. L. Wiltshire, Phys. Rev. D~{\bf 50} (1994) 7260.

\bibitem{PT}S. J. Poletti, J. Twamley, and D. L. Wiltshire, Phys. Rev. D~{\bf 51} 
(1995) 5720. 

\bibitem{PK} P. Kanti, N. E. Mavromatos, J. Rizos, K. Tamvakis, and 
E. Winstanley, Phys. Rev. D~{\bf 54} (1996) 5049.

\bibitem{TYM} T. Torii, H. Yajima and K. Maeda, Phys. Rev. D~{\bf 55} (1997) 739.

\bibitem{MSTN} K. Maeda, T. Torii, S. Suzuki, and M. Narita, in preparation.

\bibitem{SK} T. Shiromizu, K. Nakao, H, Kodama, and K, Maeda, Phys. Rev. D~{\bf 47} 
(1993) R3099.

\bibitem{M} K. Maeda, T. Koike, M. Narita, and A. Ishibashi, Phys. Rev. D~{\bf 57} 
(1998) 3503.

\bibitem{OT} It is enough to consider the case of the power law function, 
$R_{,u}\sim u^{-\alpha-1}$ to show our theorem because it gives the mildest dumping. 

\bibitem{Affin} The neighborhood of CEH can be covered by Gaussian null 
coordinates, 
$ds^2= - 2dr d\eta + f(d\eta)^2 + R^2 d\Omega$, where $r$ is an affine parameter 
of outgoing null geodesics and $f=0$ on the CEH. 
Then, $\epsilon=dr=\mbox{const.}$ 

\bibitem{MTN} It seems reasonable to assume that the surface gravity of CEH, 
$\kappa_c\equiv -f_{,r}/2$, is asymptotically constant. 
In this case, $e^{\kappa_c \eta} \sim u$ and hence $d\eta\sim du/u$.
This indicates that $dr \sim udV$. 

\bibitem{MK}
One can show that $R_{,u}<0$ on $T_C$ for large $u$ by using Eq.~(\ref{eq-R}). 
Therefore, $R_{,u}<0$ inside $T_C$ because each expansion of ingoing null 
geodesics from $T_C$ must decrease monotonically. 
Thus, we can easily get $h_{,V}>0$ by considering each segment; 
$[V_C-\epsilon/u_I,\,V_C-\epsilon/u_J]\notin \cal {U_C}$,\, 
$[V_C-\epsilon/u_J,\,V_C]\in \cal {U_C}$. The detailed proof 
can be found in Ref.~[12]. 


\end{thebibliography}
\end{document}